\documentclass[traditabstract]{aa} 
\usepackage{graphicx}
\usepackage{natbib}
\usepackage{txfonts}
\begin{document}
   \title{Mode visibilities in radial velocity and photometric Sun-as-a-star helioseismic observations}

   \author{D. Salabert
          \inst{1,2,3}\thanks{salabert@oca.eu} 
          \and
           J. Ballot
          \inst{4} 
          \and
          R.~A. Garc\'ia
          \inst{5}
          }

   \institute{Instituto de Astrof\'isica de Canarias, E-38200 La Laguna, Tenerife, Spain
         \and
          	Departamento de Astrof\'isica,  Universidad de La Laguna, E-38206 La Laguna, Tenerife. Spain
         \and
            Universit\'e de Nice Sophia-Antipolis, CNRS, Observatoire de la C\^ote dÕAzur, BP 4229, 06304 Nice Cedex 4, France
         \and
            Laboratoire d'Astrophysique de Toulouse-Tarbes, Universit\'e de Toulouse, CNRS,  31400 Toulouse, France
         \and
             Laboratoire AIM, CEA/DSM-CNRS, Universit\'e Paris 7 Diderot, IRFU/SAp, Centre de Saclay, 91191 Gif-sur-Yvette, France
             }

   \date{Received xx xx xx; accepted xx xx xx}

 
  \abstract{
 We analyze more than 5000 days of high-quality Sun-as-a-star, radial velocity GOLF and photometric VIRGO/SPM helioseismic observations to extract precise estimates of the visibilities of the low-degree p modes and the $m$-height ratios of the $l=2$ and 3 multiplets  in the solar acoustic spectrum. The mode visibilities are shown to be larger during the GOLF red-wing configuration than during the blue-wing configuration, and to decrease as the  wavelength of the VIRGO/SPM channels increases. We also show that the mode visibilities are independent of the solar activity cycle and remain constant overall with time, but that nevertheless they follow short-term fluctuations on a time scale of a few months. The $l=1$ mode visibility also increases significantly toward the end of the year 1999. Comparisons with theoretical predictions are provided. Even though there is qualitative agreement, some significant discrepancies appear, especially for the $l=3$ modes. The limb darkening alone cannot explain the relative visibilities of modes. These precise estimates should be used as references for the extraction of the p-mode parameters for any future investigation using the GOLF and VIRGO/SPM observations.
 }


  \titlerunning{Mode visibilities in Sun-as-a-star helioseismic observations}
 
   \maketitle
%

\section{Introduction}
The potential of helio- and asteroseismology to provide insights into the interiors of the Sun and other distant stars is unique. Measurements of the properties of the normal modes of solar oscillations have contributed greatly to our knowledge of both the internal structure \citep[e.g.][]{basu96,antia98,couvidat03} and dynamics \citep[e.g.][]{thompson96,chaplin99,garcia04} of the Sun's convective and radiative layers as far down to its core \citep[e.g.][]{mathur07,garcia07,mathur08}. Moreover, strong constraints on the stellar properties have also been obtained with seismic analysis from both ground-based \citep[e.g.][]{bazot05} and space-based \citep[e.g.][]{brunt09,jcd10,metcalfe10} observations.

However, our knowledge of the properties of the solar and stellar interiors depends on our ability to correctly measure the resonant modes of oscillations, which are commonly modeled as Lorentzian profiles in the acoustic power spectrum. The mode parameters (amplitude, linewidth, frequency, rotational splitting, and asymmetry) are then extracted by fitting the model to the data using maximum-likelihood functions  \citep{app98}.
The rotation of a star lifts the degeneracy of the oscillation modes and splits the eigenfrequencies into $2l+1$ $m$-components, as $-l \le m \le +l$, where $l$ is the angular degree and $m$ the azimuthal order.  In Sun-as-a-star helioseismology, only the $l+|m|$ even components are visible and it is common practice to fix the height ratios between the $m$-components of the $l = 2$ and 3 multiplets (the so-called $m$-height ratio) during the peak-fitting procedure when estimating the p-mode characteristics \citep[e.g.,][]{salabert04}, while the acoustic powers of the $l=1$, 2, and 3 modes relative to the $l=0$ modes are allowed to vary (the so-called mode visibility). However, in asteroseismology, the mode visibilities are fixed to theoretical values \citep{jcd82,palle89a} because of the lower signal-to-noise ratio (SNR) of typical data sets and shorter time series, and the $m$-height ratios are expressed as a function of the inclination of the rotation axis only \citep{app08,garcia09}. Although, the acoustic 
 frequencies \citep[e.g.,][]{woodard85}, powers, and damping rates \citep[e.g.,][]{palle90} vary with the magnetic activity cycle of the Sun and the first observation of similar variations of the p-mode parameters in a solar-like star was recently reported by \citet{garcia10}, the mode visibilities are supposed to be independent of both solar and stellar magnetic activity. \cite{froh97} observed variations with the height in the solar atmosphere  in the intensity data collected by the Variability of Solar IRradiance and Gravity Oscillation \citep[VIRGO;][]{froh95} at the beginning of the {\it Solar and Heliospheric Observatory} (SoHO) mission. However, it remains to be verified if these values are different in the radial velocity Global Oscillations at Low Frequency \citep[GOLF;][]{gabriel95} measurements between the blue- and red-wing observing periods. 

The space-based Convection, Rotation, and planetary Transits \citep[CoRoT, ][]{michel08} and Kepler \citep{borucki09} missions are currently providing in large quantities high-quality photometric measurements. After several years of Kepler observations of the same asteroseismic targets, the SNR will be high enough to measure the visibilities and $m$-height ratios in a wide range of solar-like stars in the Hertzsprung-Russell diagram at different evolution stages \citep{bedding10,chaplin10}. The forthcoming observations from the radial velocity Stellar Observations Network Group \citep[SONG, ][]{grundhal07} and the photometric PLAnetary Transits and Oscillations of stars (PLATO\footnote{\tt http://sci.esa.int/plato}) instruments will also provide extremely precise measurements of stellar oscillations in the very near future. Simultaneous radial velocity and photometric observations will be extremely useful in improving our understanding and modeling of stellar atmospheres.

   \begin{figure*}
   \centering
   \includegraphics[width=0.5\textwidth]{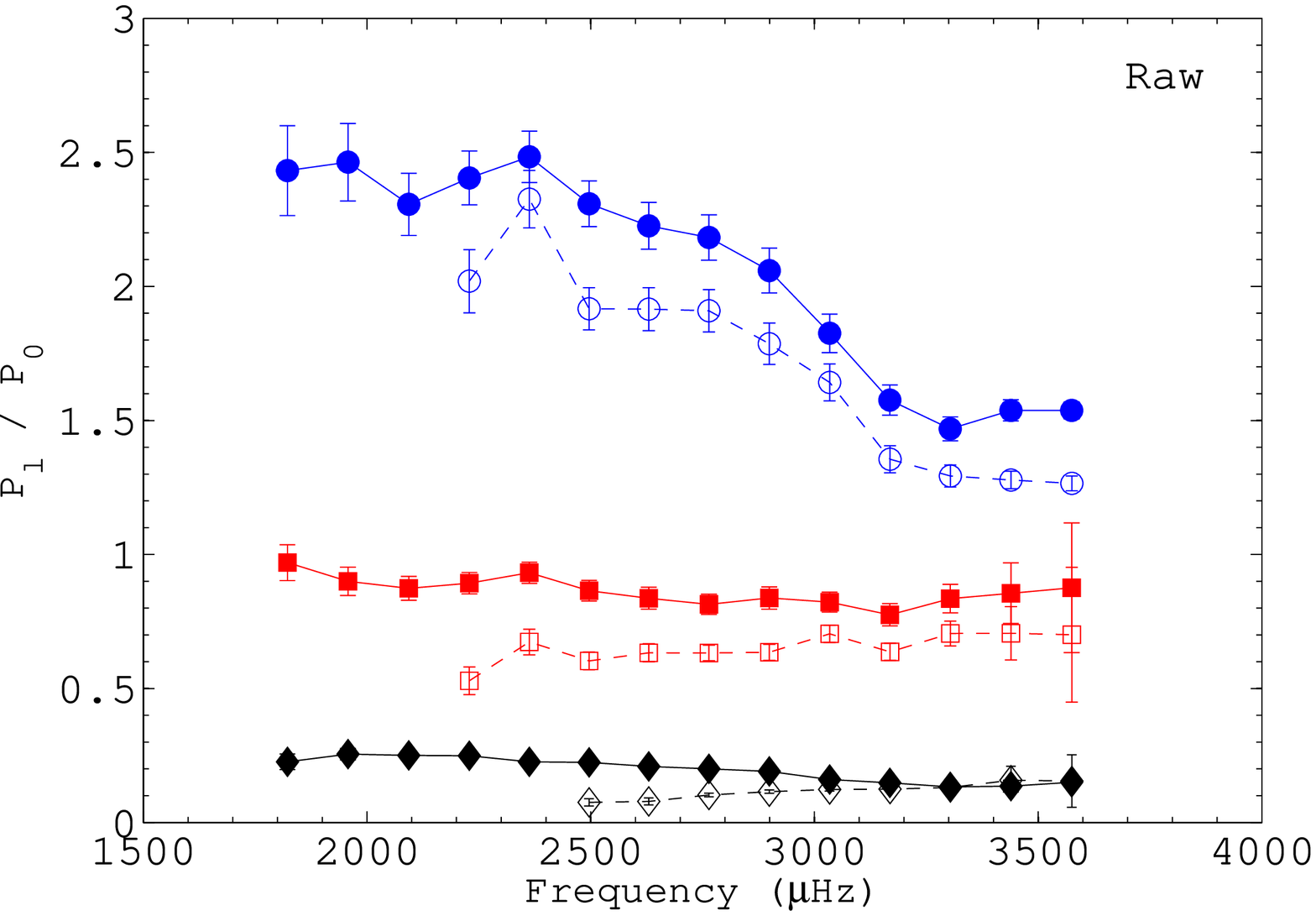}\includegraphics[width=0.5\textwidth]{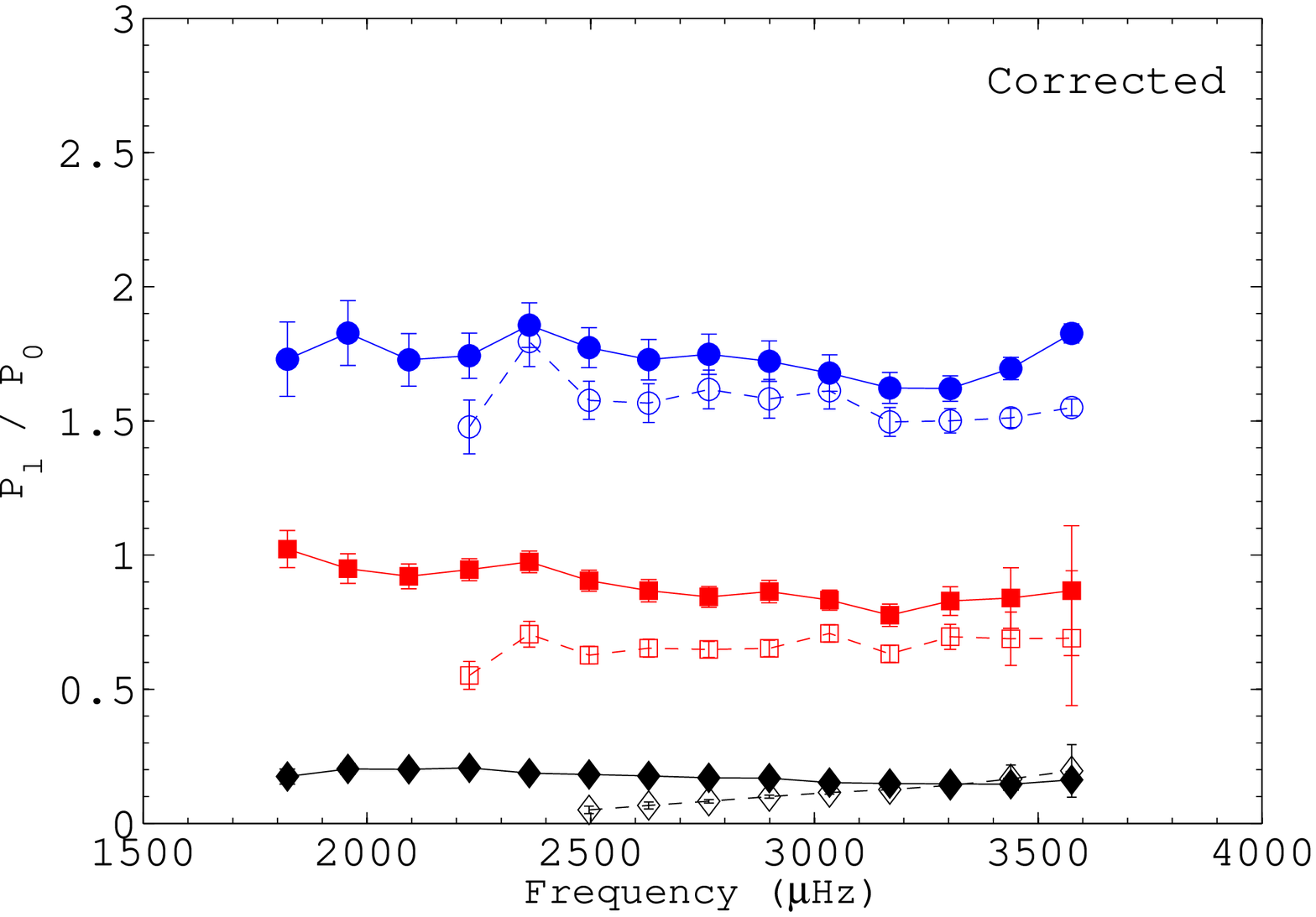}
      \caption{Raw (left) and corrected (right) mode visibilities $(V_l/V_0)^2$
 of $l = 1$ (blue dots), $l = 2$ (red squares), and $l = 3$ (black diamonds) relative to $l = 0$ as a function of frequency in the case of the radial velocity GOLF (solid lines) and intensity VIRGO/SPM (dashed lines) observations. }
         \label{fig:fig1}
   \end{figure*}
%

\section{Data and analysis}
\label{sec:data}
We analyzed simultaneous observations collected by the space-based, Sun-as-a-star Global Oscillations at Low Frequency \citep[GOLF;][]{gabriel95}  and Variability of Solar IRradiance and Gravity Oscillation  \citep[VIRGO;][]{froh95} instruments onboard the {\it Solar and Heliospheric Observatory} (SoHO) spacecraft. The GOLF instrument is a resonant scattering spectrophotometer measuring the Doppler wavelength shift -- integrated over the solar surface -- in the D$_1$ and D$_2$ Fraunhofer sodium lines at 589.6 and 589.0~nm, respectively. The VIRGO instrument is composed of three Sun photometers (SPM) at 402~nm (blue channel), 500~nm (green channel), and 862~nm (red channel). A total of 5021~days of radial velocity GOLF and intensity VIRGO/SPM observations starting on 1996 April 11 and ending on 2010 January 8 were analyzed, with respective duty cycles of  95.4\% and 94.7\%. The GOLF velocity time series were obtained following \citet{garcia05} and calibrated as described in \citet{chano03}. Moreover, in the context of this paper, it is important to remember that the GOLF signal has a small intensity pollution of around 14$\%$ \citep{palle99}. For technical reasons, GOLF has been observing only one side of the sodium doublet: in the blue wing from April 11 1996 until June 25 1998 and  again from November 18 2002 until now; and, in the meanwhile, between September 1998 until November 18 2002, the measurements were obtained from the red wing of the sodium doublet, which originates higher up in the solar atmosphere \citep{garcia05}. In addition, two extended gaps are present in the analyzed time series because of the temporary loss of the SoHO spacecraft. The first gap of $\sim$100 days happened during the summer of 1998 and was due to a failure in the gyroscopes. The second one of a period of $\sim$1 month in January 1999 occurred while new software was being uploaded to the spacecraft. 

The power spectrum of each time series was fitted to estimate the mode parameters of the $l=0$, 1, 2, and 3 modes using a multi-step iterative method \citep{salabert07}. Each mode component of radial order $n$ and angular degree $l$ with azimuthal order $m$ was parameterized with an asymmetric Lorentzian profile \citep{nigam98}, as

\begin{equation}
{\cal L}_{l,m,n}(\nu) = H_{l,m,n} \frac{(1+\alpha_{l,n} x_{l,m,n})^2+\alpha_{l,n}^2}{1+ x_{l,m,n}^2},
\label{eq:mlemodel}
\end{equation}

\noindent
where $x_{l,m,n} = 2(\nu-\nu_{l,m,n})/\Gamma_{l,n}$, and $\nu_{l,m,n}$, $\Gamma_{l,n}$, and $H_{l,m,n}$ represent the mode frequency, the linewidth, and the height of the spectral density, respectively. The peak asymmetry is described by the parameter $\alpha_{l,n}$.
The rotation is treated using a first-order approximation \citep{Ledoux51}, thus the mode frequency is expressed as $\nu_{l,m,n}=\nu_{l,n}-m\delta\nu_{l,n}$, where $\nu_{l,n}$ is the central frequency and $\delta\nu_{l,n}$ is the rotational splitting.
We note that we assumed the same linewidth for the components of a given multiplet.
Because of their close proximity in frequency, modes are fitted in pairs (i.e., $l=2$ with 0, and $l=3$ with 1). While each mode parameter within a pair of modes are free, the peak asymmetry is set to be the same within pairs of modes. When visible in the power spectra, the $l = 4$ and 5 modes were included in the fitted model. An additive value $B$ is added to the fitted profile representing a constant background noise in the fitted window. Since SoHO observes the Sun equatorwards, only the $l+|m|$ even components are visible in Sun-as-a-star observations of GOLF and VIRGO/SPM. To reduce the number of free parameters and stabilize the peak-fitting procedure, it is common practice to fix the height ratios between the visible $m$-components for both the $l=2$ and $l=3$ multiplets to their estimated values. To derive observational estimates of the $m$-height ratios, the $m=\pm2$ and $m=0$ components of the $l=2$ multiplets, and the $m=\pm3$ and $m=\pm1$ components of the $l=3$ multiplets were fitted using independent heights $H_{l,m,n}$, 
assuming that components with opposite azimuthal order $m$ have the same heights, i.e., $H_{l,-m,n}=H_{l,+m,n}$. Finally, the mode parameters were extracted by maximizing the likelihood function by assuming the power spectrum statistics is given by a $\chi^2$ with two degrees of freedom.
The natural logarithms of the mode height, linewidth, and background noise were varied resulting in normal distributions. The formal uncertainties in each parameter were then derived from the inverse Hessian matrix. The blue and red periods of GOLF were also analyzed separately, as well as the mean power spectrum of the three VIRGO/SPMs (blue, green, and red channels).

\section{Mode visibilities}
\label{sec:visi}
Because of averaging effects over the solar disk with Sun-as-a-star observations, the observed relative amplitudes of the modes are modified by visibility factors $V_l$. It is possible to recover the visibilities by assuming that the intrinsic amplitudes of modes with close frequencies are similar. By denoting $A_{l,n}$ to be the observed amplitude of a mode, we find that $V_l/V_0\approx A_{l,n}/A_{l=0,n'}$ when the modes $(l,n)$ and $(l=0,n')$ have close frequencies.

Thus, we can measure the visibilities in the spectrum as
\begin{equation}
\label{eq:v1}
(V_{1}/V_{0})^2 = P_{1,n}/P_{0,n},
\end{equation}
\begin{equation}
\label{eq:v2}
(V_{2}/V_{0})^2 = P_{2,n-1}/P_{0,n},
\end{equation}
\noindent
and
\begin{equation}
\label{eq:v3}
(V_{3}/V_{0})^2 = P_{3,n-1}/P_{0,n},
\end{equation}
\noindent
where $P_{l,n}$ is the mode power, directly linked to the square of the mode amplitude, defined as $P_{l,n} \propto H_{l,n} \Gamma_{l,n}$, while the height $H_{l,n}$ of a given multiplet $(l,n)$ is defined as the sum of the heights of its $m$-components, as: 
\begin{equation}
H_{l,n} = \sum_{m=-l}^{m=+l} H_{l,m,n}.
\end{equation}

\begin{figure}
   \centering
   \includegraphics[width=0.5\textwidth]{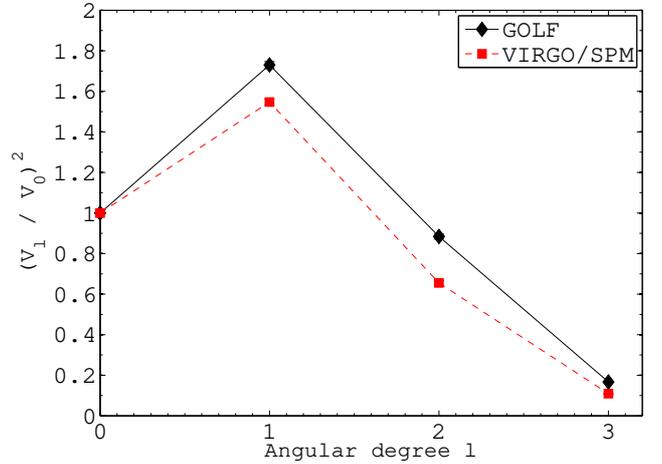}
      \caption{Mode visibilities $(V_l/V_0)^2$ averaged over frequency as a function of angular degree $l$ for the radial velocity GOLF  (black diamonds, solid line) and intensity VIRGO/SPM  (red squares, dashed line) measurements.}
         \label{fig:fig2}
   \end{figure}

\noindent
Since $\Gamma_{l,n}$ is assumed to be a smoothly varying function also of the frequency, we could directly use the ratios of the mode heights $H_{l,n}$ to measure the visibility.
We decided to use the mode powers instead because, as demonstrated by \citet{ballot08}, they can be accurately measured even when the individual $m$-components within a given multiplet start blending around 3100~$\mu$Hz because of their increasing linewidths with frequency. 
Although consistent values are obtained between the two definitions, the computation of the visibilities defined as the ratios of the mode heights is limited to frequencies lower than 3100~$\mu$Hz, as shown by \citet{salabert11}. 
The $l=1$, 2, and 3 mode visibilities calculated as in Eqs.~\ref{eq:v1}, \ref{eq:v2}, and \ref{eq:v3} are represented in Fig.~\ref{fig:fig1} as a function of frequency, for both radial velocity GOLF and intensity VIRGO/SPM (i.e., the mean power spectrum of the three SPMs on VIRGO) observations. These raw mode visibilities  (left panel of Fig.~\ref{fig:fig1}) vary with frequency -- especially for the $l=1$ mode -- that is mainly due to the 
 variation in the mode heights and linewidths with frequency. The heights and linewidths vary 
sensitively with frequency over half the large frequency separation ($\sim 67 \mu$Hz), i.e., the separation in frequency between the $l=0$ and $l=1$ modes of the same radial order $n$. Thus, the visibilities are biased  and to take account of this effect, this frequency dependence was corrected by interpolating with a spline function the powers $P_{l,n}$ of the $l=1$, 2, and 3 modes to the frequencies of the $l=0$ mode. The corrected mode visibilities are represented in the right panel of Fig.~\ref{fig:fig1}.
As the corrected mode visibilities calculated using the mode powers do not display significant variations with frequency, we can safely average them over frequency to obtain estimates of the true values. 
We note that we use the expression mode visibility here to refer to the corrected mode visibility.
Figure~\ref{fig:fig2} shows the mode visibilities $(V_1/V_0)^2$, $(V_2/V_0)^2$, and $(V_3/V_0)^2$ 
averaged over frequency, using the errors as weights, measured from GOLF and VIRGO/SPM observations as a function of the angular degree $l$. The represented values are tabulated in Table~\ref{table:visi}. The GOLF visibilities were averaged between 1800 and 3600 $\mu$Hz for all $l$ values, and the VIRGO/SPM visibilities from 2200 to 3600 $\mu$Hz for the $l=1$ and 2 modes (and from 2500~$\mu$Hz for the $l=3$ mode). The lower limits are defined by the SNR of the modes, the intensity measurements such as VIRGO/SPM having a higher noise level at low frequency and less sensitivity to the modes.
The mode visibilities in the blue-wing and the red-wing GOLF configurations, as well as the visibilities measured  separately in the blue, green, and red VIRGO/SPM channels are also given in Table~\ref{table:visi}. Significant differences are observed depending on the observing wing configuration of GOLF and the VIRGO/SPM channels analyzed. The mode visibilities are greater during the GOLF red-wing configuration than during the blue-wing configuration, the $l=3$ mode being for instance twice as  visible during the red-wing period than in the blue-wing period. The mode visibilities are shown to decrease as the wavelength of the VIRGO/SPM channels increases.

\begin{table*}[t] 
\caption{Mode visibilities in the radial velocity GOLF and intensity VIRGO/SPM measurements.}            
\label{table:visi}      
\centering                       
\begin{tabular}{l c c c c c c c}        
\hline\hline                
 $(V_l/V_0)^2$ &  GOLF  & GOLF  & GOLF &VIRGO/SPM  & VIRGO/SPM & VIRGO/SPM & VIRGO/SPM\\    
 &               &  Blue wing & Red wing & & Blue & Green & Red\\
\hline                        
$(V_1/V_0)^2 $& 1.730~$\pm$~0.017 & 1.707~$\pm$~0.023 & 1.752~$\pm$~0.028  &1.547~$\pm$~0.016  & 1.578~$\pm$~0.016 & 1.540~$\pm$~0.016 & 1.354~$\pm$~0.014\\    
$(V_2/V_0)^2$  & 0.884~$\pm$~0.012 & 0.827~$\pm$~0.017 & 1.016~$\pm$~0.022 & 0.655~$\pm$~0.012  & 0.696~$\pm$~0.012 & 0.632~$\pm$~0.011 & 0.465~$\pm$~0.008 \\
$(V_3/V_0)^2$  & 0.167~$\pm$~0.002 & 0.138~$\pm$~0.003 & 0.259~$\pm$~0.006 & 0.109~$\pm$~0.003  & 0.122~$\pm$~0.003 & 0.094~$\pm$~0.002 & 0.063~$\pm$~0.002\\
 \hline                                
\end{tabular}
\tablefoot{The mode visibilities in the blue-wing and the red-wing GOLF configurations are also given, as well as the visibilities measured separately in the blue, green, and red VIRGO/SPM channels.}
\end{table*}

%

\section{Height ratios between $m$-components}
\label{sec:mratio}
The height ratios $\beta_{l,n}$ between the visible $m$-components of the $l=2$ and $l=3$ multiplets in Sun-as-a-star observations are defined, respectively, as 
\begin{equation}
\beta_{2,n} = H_{l=2,m=0,n}/H_{l=2,m=\pm2,n},
\end{equation}

\noindent
and 
\begin{equation}
\beta_{3,n} = H_{l=3,m=\pm1,n}/H_{l=3,m=\pm3,n}.
\end{equation}

\begin{figure}
   \centering
   \includegraphics[width=0.5\textwidth]{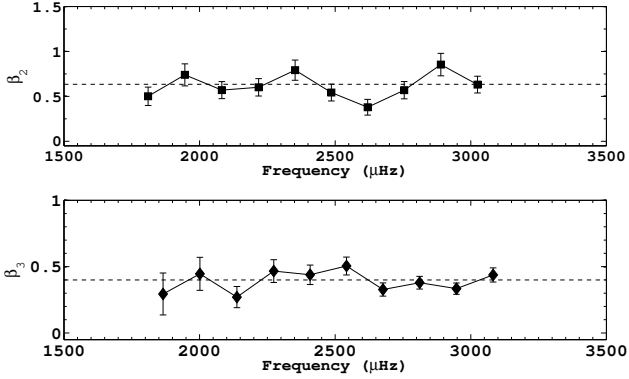}
      \caption{$m$-height ratios $\beta_l$ of the $l=2$ (top) and $l=3$ modes (bottom) as a function of frequency extracted from the GOLF radial velocity measurements.}
         \label{fig:fig3}
   \end{figure}

\noindent
As the linewidths are forced to be the same between $m$-components of a given multiplet ($l,n$) in the peak-fitting procedure, the ratios  thus measured using the heights $H_{l,m,n}$ are identical to the ones calculated using the powers $P_{l,m,n}$. These ratios $\beta_{l,n}$  are represented in Fig.~\ref{fig:fig3} as a function of frequency in the case of the GOLF observations. 
Some marginal variations with frequency can be seen in Fig.~\ref{fig:fig3}, but we assume that the height ratios are independent of frequency, and that an average over the spectrum will yield an estimate of the true value. The $m$-height ratios of the $l=2$ and $l=3$ multiplets averaged over frequency, using the errors as weights, are given in Table~\ref{table:mratio} for GOLF and VIRGO/SPM observations. No significant differences are found in the $m$-height ratios between the GOLF blue and red observing wings, and also between the three VIRGO/SPM channels. 

%

\begin{table}
\caption{Height ratios $\beta_l$ between the $m$-components of the $l=2$ and $l=3$ modes  measured in the radial velocity GOLF and intensity VIRGO/SPM observations.}            
\label{table:mratio}      
\centering                        
\begin{tabular}{lcc}        
\hline\hline                
$\beta_l$   &  GOLF  & VIRGO/SPM  \\    
\hline                        
$\beta_2$  & 0.634~$\pm$~0.033 & 0.751~$\pm$~0.059\\     
$\beta_3$  & 0.400~$\pm$~0.020 & 0.633~$\pm$~0.065     \\
\hline                                
\end{tabular}
\end{table}

\section{Modeling visibilities}

\subsection{Photometric observations}
As previously mentioned, the apparent amplitude value of a mode depends on the way that the contributions of each element of the solar disk to the total signal are averaged.
When the contribution of a solar-disk element to the total flux depends only on its distance to the limb, the mode visibility $V_l$ and the $m$-height ratio are decoupled \citep[e.g.,][]{gizon03,ballot06}. The observed amplitude $A_{n,l,m}^{\mathrm{(obs)}}$ of a mode $(n,l,m)$ is linked to the intrinsic amplitude $A_{n,l,m}$ by means of the relation
\begin{equation}
A_{n,l,m}^{\mathrm{(obs)}} = \alpha_{l,m} V_l A_{n,l,m},
\end{equation}
where
\begin{equation}
V_l = \sqrt{2\pi(2l+1)}\int_0^1 W(\mu) {\cal P}_l(\mu) \mu d\mu,\label{eq:vistheo}
\end{equation}
and
\begin{equation}
\alpha_{l,m}^2 = \frac{(l-|m|)!}{(l+|m|)!}[{\cal P}_l^{|m|}(\cos i)]^2,
\end{equation}
where ${\cal P}_l$ denote the Legendre polynomials, ${\cal P}_l^m$ are the associated Legendre functions, and $i$ is the inclination of the rotation axis relative to the line of sight, which is $i= 90\degr$ for the Sun to a very good approximation. The function $W(\mu)$ is the weighting function, which depends only on the distance to the limb $\mu$ classically defined as the cosine of the angle $\widehat{MCO}$, where $M$ is the considered point at the solar surface, $C$ the center of the Sun, and $O$ the observer.
Within this approximation, and by assuming that there is an equipartition of the energy between modes of close frequency and that the mode width depends only on the frequency, we find that
\begin{equation}
 H_{l}/H_0 =P_{l}/P_0=(V_l/V_0)^2,\mbox{ and}
\end{equation}
\begin{equation}
P_{l,m'}/P_{l,m}=H_{l,m'}/H_{l,m}=(\alpha_{l,m'}/\alpha_{l,m})^2.
\end{equation}
For photometric measurements such as VIRGO/SPM observations, the initial assumption -- $W$ depends on $\mu$ only -- is verified because the contribution depends mainly on the limb darkening. As a consequence, the amplitude ratios between the $m$-components are purely geometrical effects, and the mode visibility $V_l$ is computed by assuming the weighting function to be $W(\mu)=I_\lambda(\mu)/I(0)$, the limb-darkening profile at the observed wavelength $\lambda$. We computed the expected visibility by considering the spectral bands of the three VIRGO/SPM channels. We used the limb-darkening law from \citet{neckel94}. Results are given in Table~\ref{tab:vistheoVIRGO}.

\begin{table}
\caption{\label{tab:vistheoVIRGO} Modeled visibilities and $m$-height ratios computed for the three VIRGO/SPM channels.}
\label{tab:vistheoVIRGO}
\begin{center}
\begin{tabular}{lcccc}
\hline \hline
Mode visibility \& & Blue & Green & Red \\
$m$-height ratio& (403$\mathrm{\mu m}$) & (501$\mathrm{\mu m}$) &  (863$\mathrm{\mu m}$) \\

\hline
$(V_1/V_0)^2$ & 1.60 & 1.54 & 1.45 \\
$(V_2/V_0)^2$ & 0.66 & 0.57 & 0.45 \\
$(V_3/V_0)^2$ & 0.06 & 0.03 & 0.01 \\

\hline
$\beta_2$ & &0.667 &  \\
$\beta_3$ & &0.600 &  \\
\hline
\end{tabular}
\tablefoot{The $m$-height ratios do not depend on the channel.}
\end{center}
\end{table}

{We compare these predictions to observations (cf. Table~\ref{table:visi}). First, the variations with color are correctly reproduced for the different modes. The computations recover the observations to within a few percent accuracy, except for the $l=3$ modes, which are systematically higher than expected. Nevertheless, we note that, even for $l=1$ and 2 modes, half of the predictions have discrepancies with the observations that are larger than 3 $\sigma$. In addition, the error bars in Table~\ref{table:visi} are purely statistical, and do not take into account methodological or systematic errors. The error bars might then be underestimated. There are several possible explanations of these differences. First, the systematically larger amplitudes observed for the $l=3$ modes could be due to the intrinsic amplitudes being larger than expected. It cannot be excluded, but is hardly justifiable from a theoretical point of view today. Secondly, the differences might come from the modeled weighting function being unable to perfectly reproduce the real one. The limb-darkening law coming directly from observations is reliable. By identifying the limb-darkening profile with the weighting function, we assume that the relative intensity fluctuations $\delta I/I$ induced by a mode are the same everywhere on the visible solar disk. However, the limb is darker because the light comes mainly from higher layers of the photosphere. With an adiabatic approximation, and knowing that the temperature $T$ varies slowly in the atmosphere, it is shown that $\delta I/I=4\delta T/T\propto \xi_r $, where $\xi_r$ is the displacement. Since the mode is evanescent in the atmosphere,
 \begin{equation}
\rho\xi_r^2\propto \exp\left[-\left(1-\omega^2/\omega_c^2\right)^{1/2} h/H\right],
\end{equation}
where $h$ is the height in the atmosphere, $H$ is the pressure scale height, and $\omega_c$ is the cut-off frequency \citep[see for instance][]{jcd03}.
As the photosphere is a thin layer, we generally assume that $\xi_r$ does not vary too much. Nevertheless, because of the strong drop in density $\rho$, it is possibly not exactly true. Moreover, non-adiabitic effects are probably not negligible. Since $V_{l=3}$ modes appear to be relatively more sensitive to the shape of $W(\mu)$ close to the limb, the observed discrepencies could be due to the approximated physics of modes in the atmosphere.

Concerning the $m$-height ratio, the value are close to the observed ones (cf. Table~\ref{table:mratio}), even if, for $l=2$ modes, the observed central component is marginally larger than predicted ($\sim 1.4\sigma$). Since this term is normally insensitive to the physics of the solar atmosphere, discrepencies can be due to differences in the intrinsic amplitudes of the axisymmetric component. Nevertheless, GOLF does not have similar systematics (see next section), meaning that these differences are not conclusive.

\subsection{GOLF observations}
For GOLF observations, the weighting function does not depend only on $\mu$, and we have to perform complete computations taking into account the instrumental response, which depends on the positions in the NaD$_1$ and D$_2$ lines that are observed. Thus, mode visibilities change according to the observed wing (blue or red), but also with the epoch of the observations: according to the position in the SoHO orbit, the Doppler shift change and the lines are probed at different heights in the wings. Details of the instrumental response can be found in either \citet{garcia98} or \citet{ulrich00}.
We numerically computed the integrals over the whole disk for the different modes through a simulated instrumental response taking into account a simplified limb-darkening of the NaD lines, the solar rotation and differential rotation, and the gravitational reddening. We first assume that the modes induce purely vertical motions.

The results of these computations are listed in Table~\ref{tab:visimodel}. First, we note that the $m$-height ratio does not depend noticeably on the considered wings and is in good agreement with the observed ratio (see~Table~\ref{table:mratio}). For the visibility, the global trend is recovered, but the difference between blue and red wings for GOLF is larger than expected: the observed value for the blue wing is significantly smaller than expected.

We also performed complementary computations to take into account the horizontal motions of the modes. Since the contribution of the horizontal motions increases when the frequency decreases and the mode degree increases \citep[e.g.][]{jcd03}, it generates a small variation in the mode visibility with frequency by a few percent. Nevertheless, it is negligible compared to the observation errors. 

\begin{table}
\caption{\label{tab:visimodel} Modeled visibilities and $m$-height ratios in radial velocity for GOLF observations.}
\begin{center}
\begin{tabular}{lccc}
\hline \hline
Mode visibility \& & GOLF   & GOLF \\
$m$-height ratio     & Blue  wing   &   Red wing \\
\hline
$(V_1/V_0)^2$ & 1.84 & 1.86\\
$(V_2/V_0)^2$ & 1.08 & 1.13\\
$(V_3/V_0)^2$ & 0.27 & 0.30\\

\hline
$\beta_2$ &  0.59 & 0.58\\
$\beta_3$ & 0.43 & 0.40\\
\hline
\end{tabular}
\end{center}
\end{table}

\section{Temporal evolution of the mode visibilities}
\label{sec:temporal}
As for the p-mode frequencies \citep[e.g.,][]{woodard85,palle89b,chaplin07,salabert09}, the acoustic power and damping parameters of the low-degree acoustic modes have been found to vary with solar activity \citep[e.g.,][]{palle90,chaplin00,salabert03,simon10}. However, it has never been demonstrated whether the mode visibilities depend or not on the Sun's magnetic activity. To verify this, we analyzed the GOLF and VIRGO/SPM observations by dividing the original time series into contiguous 365-day and 91.25-day  non-independent subseries (shifted by 1/4) and fitting the power spectrum of each subseries. For the SoHO vacation in 1998, the subseries with duty cycles less than 50\% were removed from the analysis. The mode parameters were extracted in the same manner as explained in Sec.~\ref{sec:data}, with the only difference that the height ratios between the visible $m$-components of the $l=2$ and 3 modes were fixed to the values determined in Sec.~\ref{sec:mratio} and given in Table~\ref{table:mratio}, in order to stabilize the fitting procedure of these short time series.
The mode visibilities obtained as in Eqs.~\ref{eq:v1}, \ref{eq:v2}, and \ref{eq:v3} for each power spectrum were corrected for their frequency dependences by interpolating the powers of the $l=1$, 2, and 3 modes to the $l=0$ frequencies as described in Sec.~\ref{sec:visi}, and their weighted averages over frequency obtained. 

\subsection{The 365-day subseries}
Figure~\ref{fig:fig4} shows the temporal evolution of the $l=1$, 2, and 3 mode visibilities measured in the non-independent 365-day GOLF and VIRGO/SPM power spectra. The mode visibilities do not vary with solar activity, and remain constant overall over the years, within the uncertainties. However, a clear dependence on  the GOLF wing configurations, as indicated by the vertical dashed lines, can be discerned, mainly in the $l=3$ mode. The GOLF visibilities increase during the red-wing observing period between the years $\sim$~1998 and $\sim$~2002, which is consistent with the results in Table~\ref{table:visi}. On the other hand, the VIRGO/SPM visibilities do not exhibit these changes. However, the VIRGO/SPM visibility of the $l=1$ mode has a relatively large increase centered on the year 2000, which is also present in the GOLF data in addition to the larger visibilities during the red-wing operation. The VIRGO/SPM $l=2$ mode visibility increases over the same period of time appearing to follow the GOLF visibility. As the level of precision of the data, the VIRGO/SPM $l=3$ visibilities do not show any changes comparable to the ones observed in $l=1$ and 2.    
The weighted mean values over time of the GOLF visibilities are $1.747\pm0.011$, $0.862\pm0.008$, and $0.171\pm0.003$ for $l=1$, 2, and 3 respectively. In the case of VIRGO/SPM, the values are $1.605\pm0.012$, $0.641\pm0.008$, and $0.089\pm0.004$, respectively. These results are in agreement with Table~\ref{table:visi}. Similar conclusions are obtained with the individual VIRGO/SPM channels.

\subsection{The 91.25-day subseries}
Figure~\ref{fig:fig5} shows the temporal evolution of the $l=1$ and 2 mode visibilities measured in the non-independent 91.25-day GOLF and VIRGO/SPM power spectra. We do not show the results for the $l=3$ mode as they have a much larger scatter in as short time series as these, because of the mode's smaller visibility. Although, Fig.~\ref{fig:fig5} confirms that the mode visibilities are independent of the solar activity, it shows fluctuations over short periods of time, of about 6 months. The same analysis performed using different lengths of subseries and amounts of shift (1/2, 1/3, and 1/4) confirms that this six-month period is not an artifact of the methodology. Moreover, the analysis of the artificial SolarFLAG data \citep{chaplin06} does not detect this periodicity. However, its origin remains unclear.  Some of these fluctuations have significant increases, common to the GOLF and VIRGO/SPM measurements. The enhancement of the $l=1$ visibility observed in the 365-day subseries could be due to a series of higher-than-average peaks between 1999 and 2001 that are observed in the 91.25-day visibilities. Inspection of the individual mode powers seems to indicate a significant reduction in the $l=0$ power during that period. Similar results are obtained with the individual VIRGO/SPM channels.
Thus, to check whether the origin of this increase in the mode visibilities is genuine and unrelated to intrinsic changes in SoHO, we performed in Appendix~\ref{sec:app1} a similar analysis using the independent observations collected by the ground-based, multi-site network Birmingham Solar Oscillations Network \citep[BiSON;][]{elsworth95} and Global Oscillation Network Group \citep[GONG;][]{harvey96}. We show in Appendix~\ref{sec:app2} that identical temporal fluctuations seen in Figs.~\ref{fig:fig4} and \ref{fig:fig5}, including the increase in the $l=1$ visibility in 2000, are also present in the BiSON and GONG observations. 

   \begin{figure*}
   \centering
   \includegraphics[width=\textwidth]{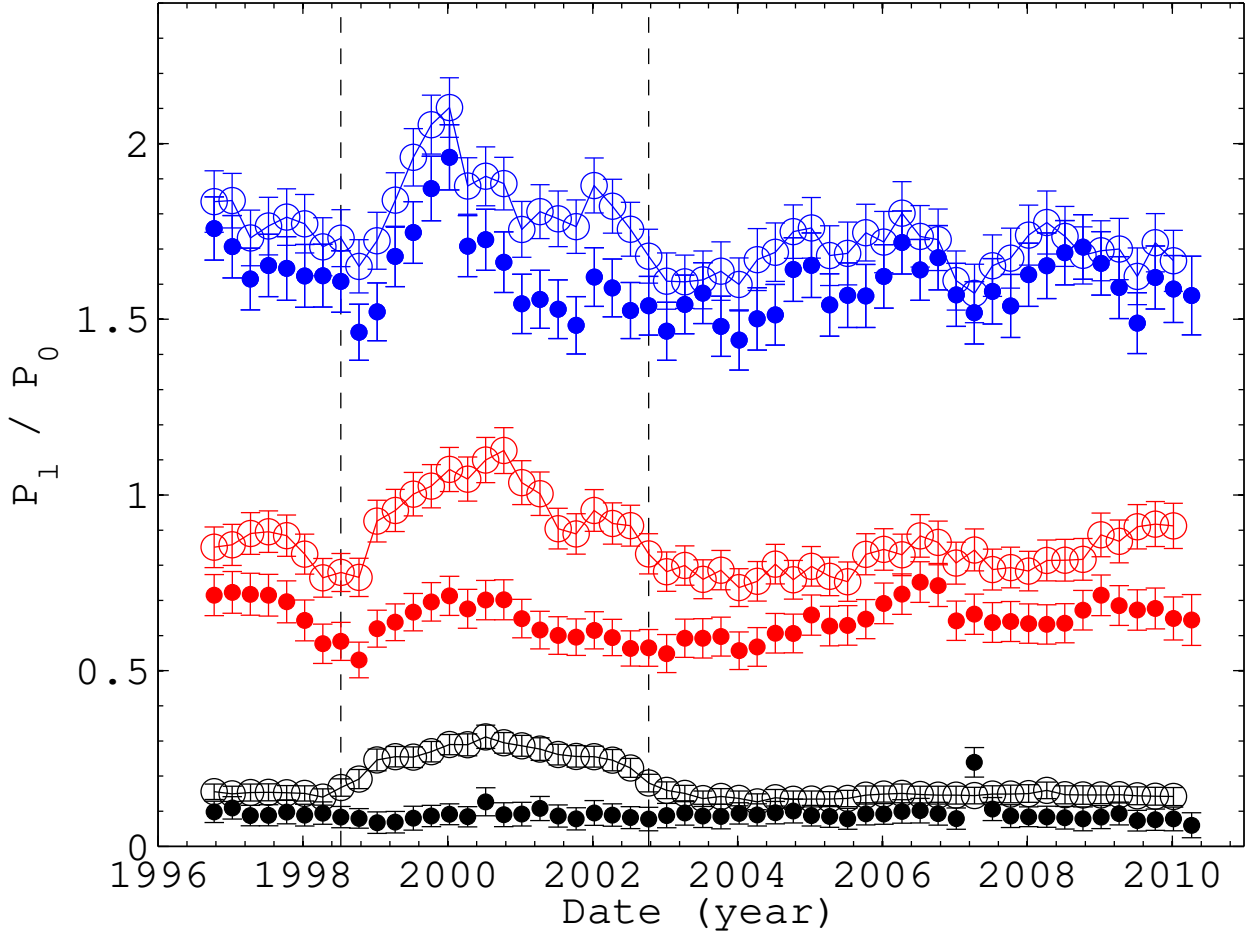} 
      \caption{Temporal evolution of the $l=1$ (blue), $l=2$ (red), and $l=3$ (black) mode visibilities measured from the analysis of the non-independent 365-day power spectra of both radial velocity GOLF (open circles) and intensity VIRGO/SPM (dots) observations. The vertical, dashed lines (from left to right) separate the periods of  blue, red, and blue-wing GOLF operations, respectively.}
         \label{fig:fig4}
   \end{figure*}
%

   \begin{figure*}
   \centering
   \includegraphics[width=\textwidth]{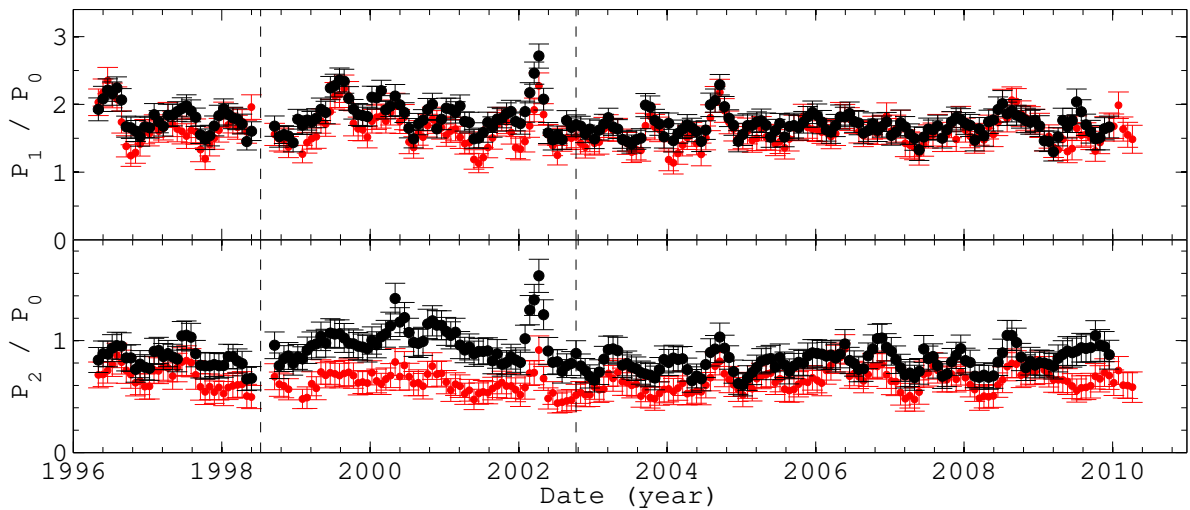}
      \caption{Temporal evolution of the $l=1$ (top) and $l=2$ (bottom) mode visibilities measured from the analysis of the non-independent 91.25-day power spectra of both radial velocity GOLF (black) and intensity VIRGO/SPM (red) observations. The vertical, dashed lines (from left to right) separate the periods of blue, red, and blue-wing GOLF operations, respectively.}
         \label{fig:fig5}
   \end{figure*}
%

\section{Conclusions}
We have analyzed more than 5000 days of high-quality helioseismic observations collected by the space-based, Sun-as-a-star, radial velocity GOLF and intensity VIRGO/SPM instruments onboard the SoHO spacecraft to extract precise measurements of the low-degree, p-mode visibilities and $m$-height ratios. The visibilities of the $l=1$, 2, and 3 modes relative to the $l=0$ mode are defined as the ratios $P_{l=1} / P_{l=0}$, $P_{l=2} / P_{l=0}$, and $P_{l=3}/ P_{l=0}$ respectively, where $P_l$ is the acoustic power of a given mode of angular degree $l$. The height ratios between the $m$-components of the $l=2$ and $l=3$ multiplets are defined as  $H_{l=2,m=0}/H_{l=2,m=\pm2}$ and $H_{l=3,m=\pm1}/H_{l=3,m=\pm3}$ respectively, where $H_{l,m}$ is the height of the individual $m$-components of the mode. We derived observational values that we then compared to theoretical values computed using the instrumental responses of GOLF and VIRGO/SPM by taking into account limb-darkening functions. In general, the predicted values qualitatively agree with the observations. We also showed that the mode visibility is a function of the wing configuration of GOLF, the $l=1$, 2, and 3 modes being more visible during the red-wing period than during the blue-wing one. The mode visibilities were also shown to depend on the wavelength of the VIRGO/SPMs, as predicted. However, some discrepancies between observations and theory remain, especially for the $l=3$ modes. These differences imply that the limb darkening alone is insufficient to explain the observed visibilities. This has to be analyzed in depth to be able to make reliable predictions for other stars. Nevertheless, we note that the discrepencies identified here are smaller than the error bars in mode amplitudes obtained with a few months of CoRoT or Kepler observations.

By analyzing 365-day and 91.25-day GOLF and VIRGO/SPM subseries, we also demonstrated for the first time that the mode visibilities are constant in time and do not vary with the solar magnetic activity. However, the mode visibilities exhibit short-term temporal variations of about six months that are common to the GOLF and VIRGO/SPM observations. Some of these fluctuations increase significantly, and a significant increase in the $l=1$ mode visibility was found at the end of the year 1999. The origin of both these short-period fluctuations and this bump at the end of 1999 is unclear but might be related to the stochastic excitation of the individual modes or the variations in the acoustic power and damping rate of the solar oscillations with time. 

In the future, the observations from the radial velocity SONG and photometric PLATO instruments will provide invaluable inputs to model the stellar atmospheres.
Finally, these newly constrained $l=1$, 2, and 3 mode visibilities and $m$-height ratios provided in this paper should be used as inputs to any investigation of the Sun using the radial velocity GOLF and intensity VIRGO/SPM observations.

\begin{acknowledgements}
The authors wish to thank Catherine Renaud and Antonio Jim\'enez for the calibration and preparation of the GOLF and VIRGO/SPM datasets. The GOLF and VIRGO/SPM instruments onboard SoHO are a cooperative effort of many individuals, to whom we are indebted. SoHO is a project of an international collaboration between ESA and NASA. DS acknowledges the support of the grant PNAyA2007-62650 from the Spanish National Research Plan. This work was supported by the CNES/GOLF grant at SAp/CEA-Saclay. This work utilizes data obtained by the Global Oscillation Network Group (GONG) program, managed by the National Solar Observatory, which is operated by AURA, Inc. under a cooperative agreement with the National Science Foundation. The data were acquired by instruments operated by the Big Bear Solar Observatory, High Altitude Observatory, Learmonth Solar Observatory, Udaipur Solar Observatory, Instituto de Astrof\'{\i}sica de Canarias, and Cerro Tololo Interamerican Observatory. This work also utilizes data collected by the Birmingham Solar Oscillations Network (BiSON), which is funded by the UK Science Technology and Facilities Council (STFC). We thank the members of the BiSON team, colleagues at the host institutes, and all others, past and present, who have been associated with BiSON. 
 \end{acknowledgements}

\begin{appendix} 

\section{Mode visibilities in BiSON and GONG integrated observations}
\label{sec:app1}

This analysis presented in these appendices was motivated by the observation of an increase in the temporal evolution of the visibilities, mainly for the $l=1$ mode, at the end of the year 1999, seen in the Sun-as-a-star GOLF and VIRGO/SPM instruments onboard the SoHO spacecraft. We wish here and in Appendix~\ref{sec:app2} to establish whether the origin of these variations might come from SoHO itself and not from the Sun. We decided then to perform the same analysis using two independent helioseismic datasets collected by the ground-based, multi-site BiSON, and GONG networks. 
The unresolved BiSON and spatially-resolved GONG networks are both composed of six stations located at selected longitudes around the world. The instruments at each BiSON\footnote{Respective datasets are available from {\tt http://bison.ph.bham.ac.uk/data.php} and {\tt http://gong.nso.edu/data/}.} site make Sun-as-a-star observations of the Doppler shift of the potassium Fraunhofer line at 770~nm \citep{elsworth95}, while the GONG$^2$ cameras use Michelson Doppler interferometer-based instruments measuring in the absorption line Ni I at 676.8~nm \citep{harvey96}. In these appendices, we used the spatially-integrated GONG time series, which is analogous to the unresolved observations of GOLF, VIRGO/SPM, and BiSON, although the visibilities for the higher degree modes are different, with the relative strength falling off faster as a function of $l$ in the GONG data. A total of 5339 days, starting on 1995 January 1, of BiSON observations, and of 5112 days, starting on 1995 May 7, of the GONG integrated data, with respective duty cycles of 81.0\% and 85.4\%, were analyzed in the same manner as described in Sec.~ \ref{sec:data}. The daily temporal sidebands due to the diurnal gaps in ground-based observations were included in the fitting model (Sec.~\ref{sec:data}). 
The BiSON and GONG integrated $l=1$, 2, and 3 mode visibilities, obtained as explained in Sec.~\ref{sec:visi}, are represented as a function of frequency on Fig.~\ref{fig:fig6}, and given in Table~\ref{table:visi_gong} once averaged over frequency between 1800 and 3100~$\mu$Hz. The $l=3$ mode has a very small visibility in the GONG integrated data and was averaged from 2500~$\mu$Hz only, due to its lower SNR at low frequency.
The $m$-height ratios were also estimated and are given in Table~\ref{table:mratio_gong}. \citet{chaplin01} calculated the $m$-height ratios in the BiSON data finding $0.55\pm0.04$ and $0.38\pm0.02$ for the $l=2$ and $l=3$ modes, respectively. Our analysis returns larger ratios than theirs, but are consistent within $1\sigma$ for $l=2$ and $2\sigma$ for $l=3$. These differences might be cause by differences in the total length of the datasets used, 8 years in \citet{chaplin01} compared to 14 years here, meaning a higher frequency resolution in the present case and different periods of time analyzed -- indeed, variations with solar activity cannot be totally excluded. We calculated these height ratios over the same period of time as in \citet{chaplin01} and found $0.57\pm0.04$ and $0.40\pm0.03$ for $l=2$ and 3, respectively, which agree with \citet{chaplin01} to $1\sigma$. The remaining differences might be attributable to the peak-fitting itself, such as the size of the fitting windows, or the small multiplet frequency asymmetry at $l=2$ and 3, which is not taken into account in our analysis.
In the case of the integrated GONG time series, no previously measured $m$-height ratios were found in the published litterature. We note that since the visibility of the $l=3$ mode is close to being zero in the GONG integrated data, the $m$-height ratio for $l=3$ could not be measured accurately.

   \begin{figure*}
   \centering
   \includegraphics[width=0.5\textwidth]{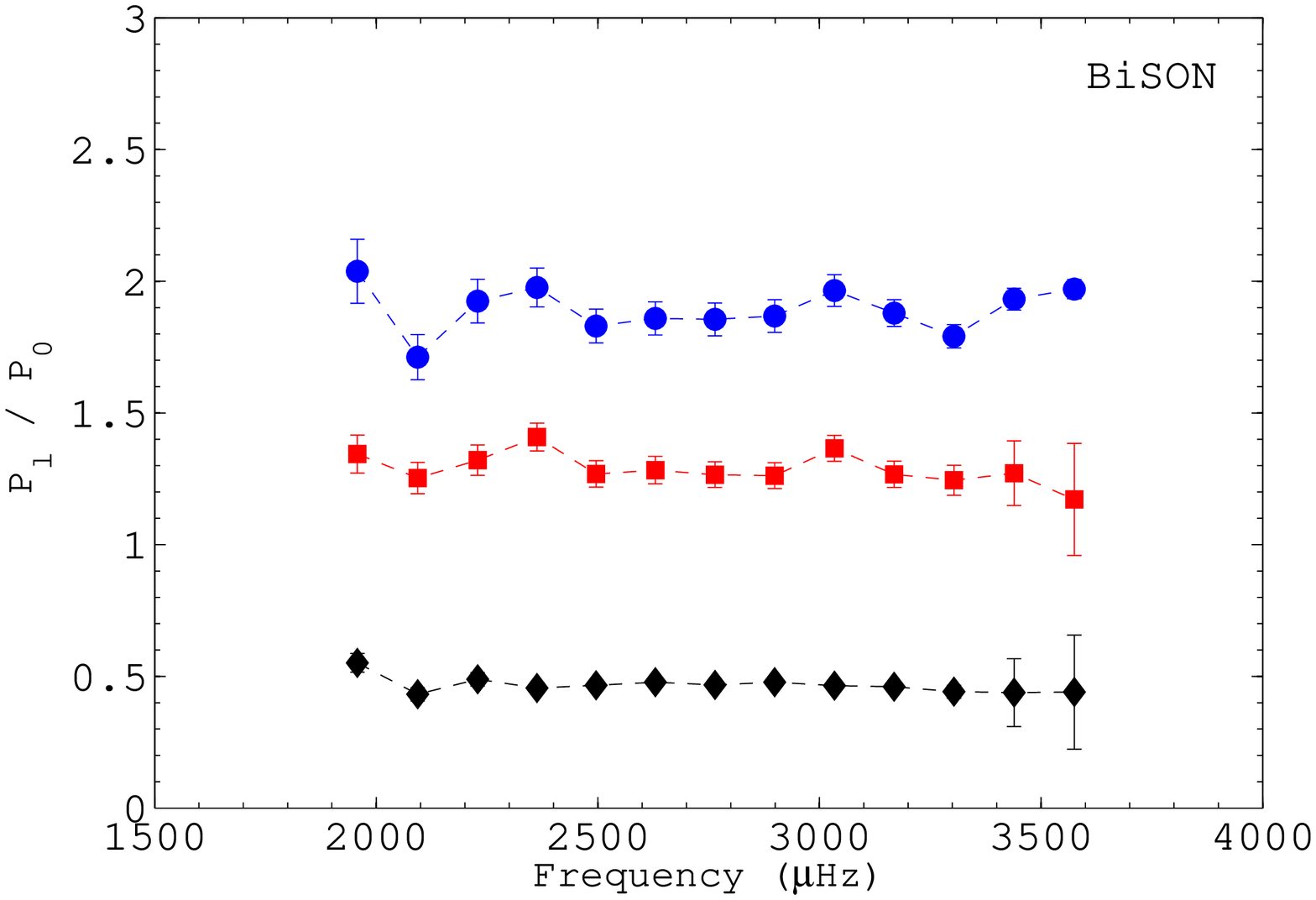}\includegraphics[width=0.5\textwidth]{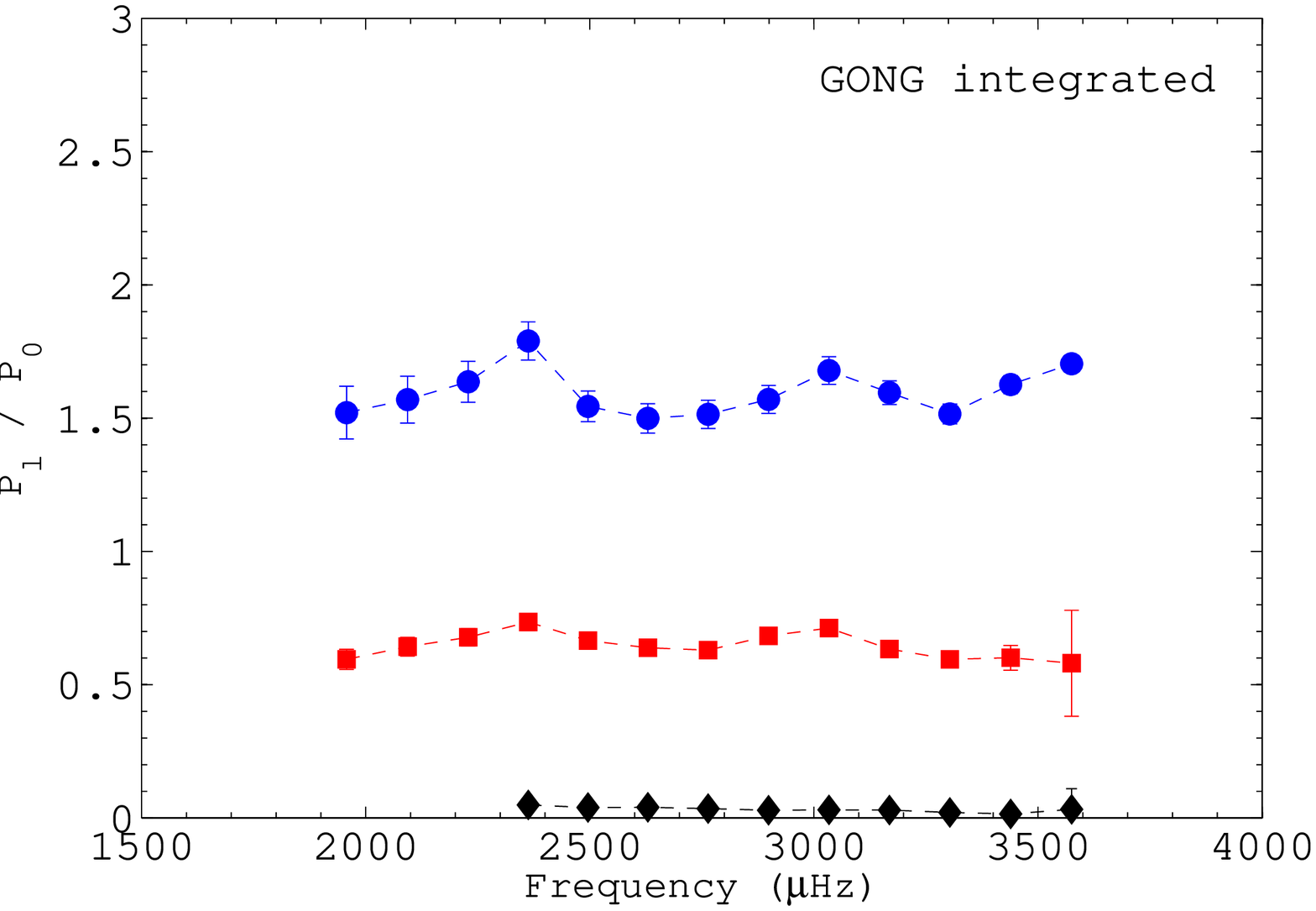}
         \caption{Mode visibilities of $l = 1$ (blue dots), $l = 2$ (red squares), and $l = 3$ (black diamonds) relative to $l = 0$ as a function of frequency extracted from the analysis of the BiSON (left) and GONG integrated (right) observations.}
         \label{fig:fig6}
   \end{figure*}
%

   \begin{figure*}
   \centering
   \includegraphics[width=0.7\textwidth,angle=90]{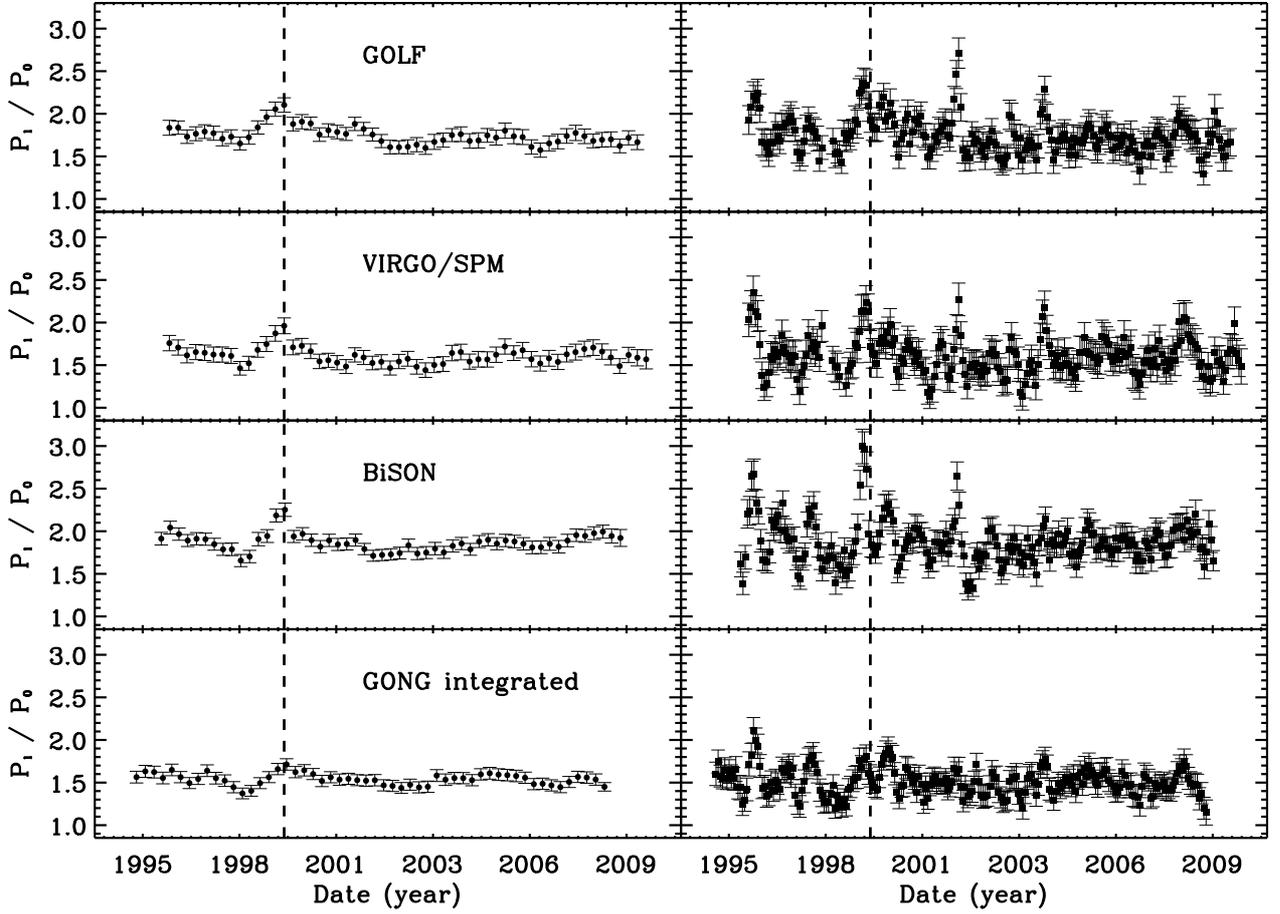}
      \caption{Temporal variations in the $l=1$ mode visibilities extracted from the analysis of the non-independent 365-day (left column) and 91.25-day (right column) power spectra of the GOLF, VIRGO/SPM, GONG, and BiSON (from top to bottom) instruments. The vertical dashed lines correspond to 1999 October 10.}
         \label{fig:fig7}
   \end{figure*}
%

\begin{table}
\caption{Mode visibilities extracted from the analysis of the BiSON and integrated GONG time series.}            
\label{table:visi_gong}      
\centering                        
\begin{tabular}{lcc}        
\hline\hline                
$(V_l/V_0)^2$   &  BiSON & GONG   \\    
\hline                        
$(V_1 / V_0)^2$ & 1.895~$\pm$~0.016 & 1.606~$\pm$~0.014\\     
$(V_2 / V_0)^2$ & 1.296~$\pm$~0.016 & 0.652~$\pm$~0.008\\
$(V_3 / V_0)^2$ & 0.467~$\pm$~0.006 & 0.028~$\pm$~0.001\\
\hline                                
\end{tabular}
\end{table}

\begin{table}
\caption{Height ratios $\beta_l$ between the $m$-components of the $l=2$ and $l=3$ modes measured in the BiSON and integrated GONG time series.}            
\label{table:mratio_gong}      
\centering                        
\begin{tabular}{lcc}        
\hline\hline                
$\beta_l$   & BiSON &  GONG   \\    
\hline                        
$\beta_2$  & 0.625~$\pm$~0.032 & 0.882~$\pm$~0.045\\     
$\beta_3$ & 0.445~$\pm$~0.020 & ---\tablefootmark{a} \\
\hline                                
\end{tabular} 
\tablefoot{
\tablefoottext{a}{Owing to the very small visibility of the $l=3$ mode in the GONG integrated data, the $m$-height ratio for $l=3$ could not be accurately measured.} 
}
\end{table}

\section{Temporal evolution of the $l=1$ mode visibility in several Sun-as-a-star helioseismic instruments}
\label{sec:app2}
We compare the temporal evolution of the low-degree mode visibilities of the solar oscillations measured with the spaced-based GOLF and VIRGO/SPM instruments onboard the SoHO spacecraft, to the visibilities measured with the ground-based, multi-site BiSON, and GONG networks. The BiSON and GONG integrated time series were divided into contiguous 365-day and 91.25-day  non-independent subseries (shifted by 1/4) and analyzed in the same manner as explained in Sec.~ \ref{sec:temporal}. 
Figure~\ref{fig:fig7} shows the temporal evolution of the visibility of the $l=1$ mode extracted from the analysis of the non-independent 365-day and 91.25-day GOLF, VIRGO/SPM, BiSON, and GONG integrated power spectra.
The bump of the $l=1$ mode visibility at the end of the year 1999 observed in the 365-day subseries (see Fig. ~\ref{fig:fig4}) is present in all the analyzed independent datasets.  As already mentioned in Sec.~\ref{sec:temporal}, this bump is most likely the result of a series of higher-than-average peaks between 1999 and 2001 observed in the 91.25-day visibilities. Moreover, similar short-term variations of about 6 months can be observed in the 91.25-days subseries of the GOLF, VIRGO/SPM, BiSON, and GONG instruments, some of them exhibiting significant and common increases. The same analysis performed using different lengths of subseries and overlapping factors confirms that this six-month periodicity is not an artifact of the methodology.

\end{appendix}

\end{document}